\pdfoutput=1
\documentclass[twocolumn,showpacs,preprintnumbers,superscriptaddress,amsmath,amssymb,aps,pre]{revtex4}
\usepackage{graphicx,subfigure}
\usepackage{dcolumn}
\usepackage{bm}
\usepackage{color}

\begin{document}

\title{Dynamic evolution of cross-correlations in the Chinese stock market}

\author{Fei Ren}
 \email{fren@ecust.edu.cn}
 \affiliation{School of Business, East China University of Science and Technology, Shanghai 200237, China} %
 \affiliation{Research Center for Econophysics, East China University of Science and Technology, Shanghai 200237, China} %
 \affiliation{School of Science, East China University of Science and Technology, Shanghai 200237, China} %

\author{Wei-Xing Zhou}
 \email{wxzhou@ecust.edu.cn}
 \affiliation{School of Business, East China University of Science and Technology, Shanghai 200237, China} %
 \affiliation{Research Center for Econophysics, East China University of Science and Technology, Shanghai 200237, China} %
 \affiliation{School of Science, East China University of Science and Technology, Shanghai 200237, China} %

\date{\today}

\begin{abstract}
We study the dynamic evolution of cross-correlations in the Chinese stock market mainly based on the random matrix theory (RMT). The correlation matrices constructed from the return series of 367 A-share stocks traded on the Shanghai Stock Exchange from January 4, 1999 to December 30, 2011 are calculated over a moving window with a size of 400 days. The evolutions of the statistical properties of the correlation coefficients, eigenvalues, and eigenvectors of the correlation matrices are carefully analyzed. We find that the stock correlations are significantly increased in the periods of two market crashes in 2001 and 2008, during which only five eigenvalues significantly deviate from the random correlation matrix, and the systemic risk is higher in these volatile periods than calm periods. By investigating the significant contributors of the deviating eigenvectors in different moving windows, we observe a dynamic evolution behavior in business sectors such as IT, electronics, and real estate, which lead the rise (drop) before (after) the crashes.
\end{abstract}


\maketitle


\section{Introduction}

The stock market is a typical complex system with interactions between individuals, groups, and institutions at different levels. In financial crises, the risk can quickly propagate among these interconnected institutions which have mutual beneficial business. Therefore, the analysis of the correlations between shares issued by different institutions is of crucial importance for the understanding of interactive mechanism of the stock market and the portfolio risk estimation \cite{Farmer-1999-CSE,Mantegna-Stanley-2000,Bouchaud-Potters-2000}. Variety of works have been done to reveal the information contained in the internal correlations among stocks, and the methods generally used in the research of stock cross-correlations include the random matrix theory (RMT) \cite{Plerou-Gopikrishnan-Rosenow-Amaral-Stanley-1999-PRL,Laloux-Cizean-Bouchaud-Potters-1999-PRL}, the principal component analysis (PCA) \cite{Zheng-Podobnik-Feng-Li-2012-SR,Billio-Getmansky-Lo-Pelizzon-2012-JFE,Kritzman-Li-Page-Rigobon-2011-JPM}, and the hierarchical structure \cite{Mantegna-1999-EPJB,Bonanno-Caldarelli-Lillo-Mantegna-2003-PRE,Bonanno-Caldarelli-Lillo-Micciche-Vandewalle-Mantegna-2004-EPJB,Tumminello-Coronnello-Lillo-Micciche-Mantegna-2007-IJBC,Tumminello-DiMatteo-Aste-Mantegna-2007-EPJB,Garas-Argyrakis-2009-EPL,Cai-Zhou-Zhou-Zhou-2010-IJMPC,Tumminello-Lillo-Mantegna-2010-JEBO,Kwapien-Drozdz-2012-PR}.

The random matrix theory (RMT), originally developed in complex quantum system, is applied to analyze the cross-correlations between stocks in the U.S. stock market by Plerou {\em{et al.}} \cite{Plerou-Gopikrishnan-Rosenow-Amaral-Stanley-1999-PRL}. The statistics of the most of the eigenvalues of the correlation matrix calculated from stock return series agree with the predictions of random matrix theory, but with deviations for a few of the largest eigenvalues. Extended work has been conducted to explain information contained in the deviating eigenvalues \cite{Plerou-Gopikrishnan-Rosenow-Amaral-Guhr-Stanley-2002-PRE}, which reveals that the largest eigenvalue corresponds to a market-wide influence to all stocks and the remaining deviating eigenvalues correspond to conventionally identified business sectors. Additional work has proved that even the eigenvalues within the spectrum of RMT carry some sort of correlations \cite{Kwapien-Drozdz-Oswiecimka-2006-pa,Drozdz-Kwapien-Oswiecimka-2007-appb}. Using the same RMT method, extensive works have been performed in the correlation analysis of various stock markets \cite{Utsugi-Ino-Oshikawa-2004-PRE,Kim-Jeong-2005-PRE,Pan-Sinha-2007-PRE,Wilcox-Gebbie-2007-PA,Shapira-Kenett-Jacob-2009-EPJB,Shen-Zheng-2009a-EPL,Ahn-Choi-Lim-Cha-Kim-Kim-2011-PA,Oh-Eom-Wang-Jung-Stanley-Kim-2011-EPJB,Wang-Podobnik-Horvatic-Stanley-2011-PRE,Kwon-Oh-2012-EPL}.

In recent years, there are increasing works concentrated on the variation of the cross-correlations between market equities over time \cite{Drozdz-Grummer-Gorski-Ruf-Speth-2000-pa,Drozdz-Grummer-Ruf-Speth-2001-pa,Onnela-Chakraborti-Kaski-Kertesz-Kanto-2003-PRE,Aste-Shaw-Matteo-2010-NJP,Podobnik-Wang-Horvatic-Grosse-Stanley-2010-EPL,Qiu-Zheng-Chen-2010-NJP,Song-Tumminello-Zhou-Mantegna-2011-PRE,Fenn-Porter-Williams-McDonald-Johnson-Jones-2011-PRE,Kumar-Deo-2012-PRE,Gao-Wei-Wang-2013-IJMPC}. Aste {\em{et al.}} have investigated the evolution of the correlation structure among 395 stocks quoted on the U.S. equity market from 1996 to 2009, in which the connected links among stocks are built by a topologically constrained graph approach \cite{Aste-Shaw-Matteo-2010-NJP}. They find that the stocks have increased correlations in the period of larger market instabilities. By using the similar filtered graph approach, the correlation structure among 57 different market indices all over the world has been studied \cite{Song-Tumminello-Zhou-Mantegna-2011-PRE}. Fenn {\em{et al.}} have used the RMT method to analyze the time evolutions of the correlations between the market equity indices of 28 geographical regions from 1999 to 2010 \cite{Fenn-Porter-Williams-McDonald-Johnson-Jones-2011-PRE}, and they also observe the increase of the correlations between several different markets after the credit crisis of 2007-2008. Similar results have also been observed in Refs. \cite{Drozdz-Grummer-Gorski-Ruf-Speth-2000-pa,Drozdz-Grummer-Ruf-Speth-2001-pa,Podobnik-Wang-Horvatic-Grosse-Stanley-2010-EPL}.

The RMT method has been applied to the analysis of the static correlations between the return series in the Chinese stock market \cite{Shen-Zheng-2009a-EPL}. No clear interactions between stocks in same business sectors are observed, while unusual sectors containing the ST (specially treated) and Blue-chip stocks are identified by a few of the largest eigenvalues. Further work has been done to analyze the anti-correlated sub-sectors that compose the unusual sectors \cite{Jiang-Zheng-2012-EPL}. Up to now, not much work has been conducted on the dynamics of stock correlations in the Chinese market to the best of our knowledge. Using the daily records of 259 stocks on the Chinese stock market from 1997 to 2007, the dynamic evolution of the Chinese stock network was firstly analyzed in ~\cite{Qiu-Zheng-Chen-2010-NJP}. In their work the links are constructed between the stocks which have correlations larger than a threshold, and a stable topological structure is revealed by using a dynamic threshold instead of the static threshold. Although additional efforts are made to identify the economic sectors based on the RMT method, the dynamic effects of conventional business sectors is extremely week.

The principal component analysis (PCA) is another method commonly used to detect the correlations between stock returns. It is closely related to the RMT method, since it is also done through eigenvalue decomposition of the correlation (or covariance) matrix of the return series. This method uses an orthogonal transformation to convert a set of possible correlated returns into several uncorrelated components, which are ranked by their explanatory power for the total variance of the system. The studies of correlations among stock returns based on the PCA method are primarily concerned about the systemic risk measures \cite{Zheng-Podobnik-Feng-Li-2012-SR,Billio-Getmansky-Lo-Pelizzon-2012-JFE,Kritzman-Li-Page-Rigobon-2011-JPM}.

In this paper, by mainly using the RMT method, we study dynamic evolution of the correlations between the 367 A-share stocks traded on Shanghai Stock Exchange from 1999 to 2011. The internal correlations between the stocks are investigated based on the correlation matrix of the return series of individual stocks in a moving window with a fixed length. We mainly concern about the statistical properties of the correlation coefficients, eigenvalues and eigenvectors of the correlation matrix, and their variations in different time periods. Our results confirm the strong collective behavior of the stock returns in the periods of market crashes, which is verified by the observations of the distribution of the correlation coefficients and the mean correlation coefficient. Further, based on the PCA method we calculate the proportion of total variance explain by the first $n$ components, through which the systemic risk of the Chinese stock market is estimated for different time periods. Another important purpose of our study is to extract the information contained in the eigenvectors deviating from RMT. We find the largest eigenvector quantifies a market-wide influence on all stocks, and this market mode remains stable over time. For the interpretations of other deviating eigenvectors, dynamic evolutions of several conventional industries including IT, electronics, machinery, petrochemicals, and real estate, are remarkably observed.

The paper is organized as follows. Section 2 offers a brief description of the data analyzed, and introduces the calculation of the correlation coefficients as the elements of the correlation matrix. Section 3 discusses the variation of the statistical properties of the correlation coefficients evolved with the historical time. In Section 4, we study the dynamic behaviors of the eigenvalues and their significance, and discuss the use of eigenvalues for the explanation of total variance. Section 5 discusses the evolutions of the statistical properties of the eigenvectors, and provides interpretations for the deviating eigenvectors. Finally, Section 6 gives the conclusion.

\section{Data and construction of correlation matrix}
The database analyzed in our study contains the daily data of all A-Share stocks traded on Shanghai Stock Exchange (SHSE), one of the two stock exchanges in mainland China. The A-Share stocks are issued by mainland Chinese companies, and traded in Chinese Yuan. The data source is from Beijing Gildata RESSET Data Technology Co., Ltd, see http://www.resset.cn/. To better understand the correlation structures under different market conditions, we select the A-share stocks traded on Shanghai Stock Exchange from January 4, 1999 to December 30, 2011 covering the two big crashes in 2001 and 2008. To make sure that the stocks have enough number of trading days to be statistically significant in our studies, we select the stocks traded on the stock exchange for at least 2600 days, i.e., exclude those stocks suspended from the market for more than about two years. This filter yields the sample data including 367 A-Share stocks and 1114364 daily records in total.

Before we quantify the cross-correlations between stocks, we first calculate the return series for a given stock $i$ as
\begin{equation}
G_i(t)=\ln p_i(t)-\ln p_i(t-1), \label{equ_return}
\end{equation}
where $p_i(t)$ is the price for stock $i$ at time $t$, and $t$ is in units of one day. The Pearson's correlation coefficient between two stock return series $G_i(t)$ and $G_j(t)$ is defined as
\begin{equation}
c_{ij}=\frac{\langle (G_i(t)- \langle G_i(t)\rangle) (G_j(t)- \langle G_j(t)\rangle) \rangle}{\sigma_i \sigma_j}, \label{equ_correlation}
\end{equation}
where $\sigma_i$ and $\sigma_j$ are the standard deviations of two stock return series. It is a common measure of the dependence between the return series of the two stocks. There are $N=367$ sample stocks, therefore we have a correlation matrix $C$ with $367\times367$ correlation coefficients as elements. The elements of the correlation matrix are restricted to the domain $-1\leq c_{ij}\leq1$: for $0 < c_{ij} \leq 1$ the stocks are correlated, for $-1 \leq c_{ij}<0$ the stocks are anti-correlated, and for $c_{ij}=0$ the stocks are uncorrelated.

The cross-correlation defined above is to calculate the dependence between the return series in the whole period of the sample data. We are more interested in the dynamic variation of the stock correlations evolved with time $t$, so we look at the correlations calculated over a moving window. The size $T$ of the moving window is fixed to be 400 trading days, i.e., about two years, which is a little bit larger than the number of the sample stocks. Equation~\ref{equ_correlation} is applied to calculate the correlation coefficients over a subset of return series within the moving window $[t-T+1,t]$. For instance, the correlations in the first moving window are computed by the return series within $[1,T]$, and $[2,T+1]$ for the following moving window. In consideration of our sample date, which is from 04/01/1999 to 30/12/2011, the starting date of the moving window covers the period from 04/01/1999 to 12/05/2010, and the ending date is from 06/09/2000 to 30/12/2011.

\section{Evolutions of statistical properties of correlation coefficients}

\subsection{Distribution of correlation coefficients}

\begin{figure}
\centering
\subfigure[][]{\label{Fig:return:PDF:correff-a}\includegraphics[width=8cm]{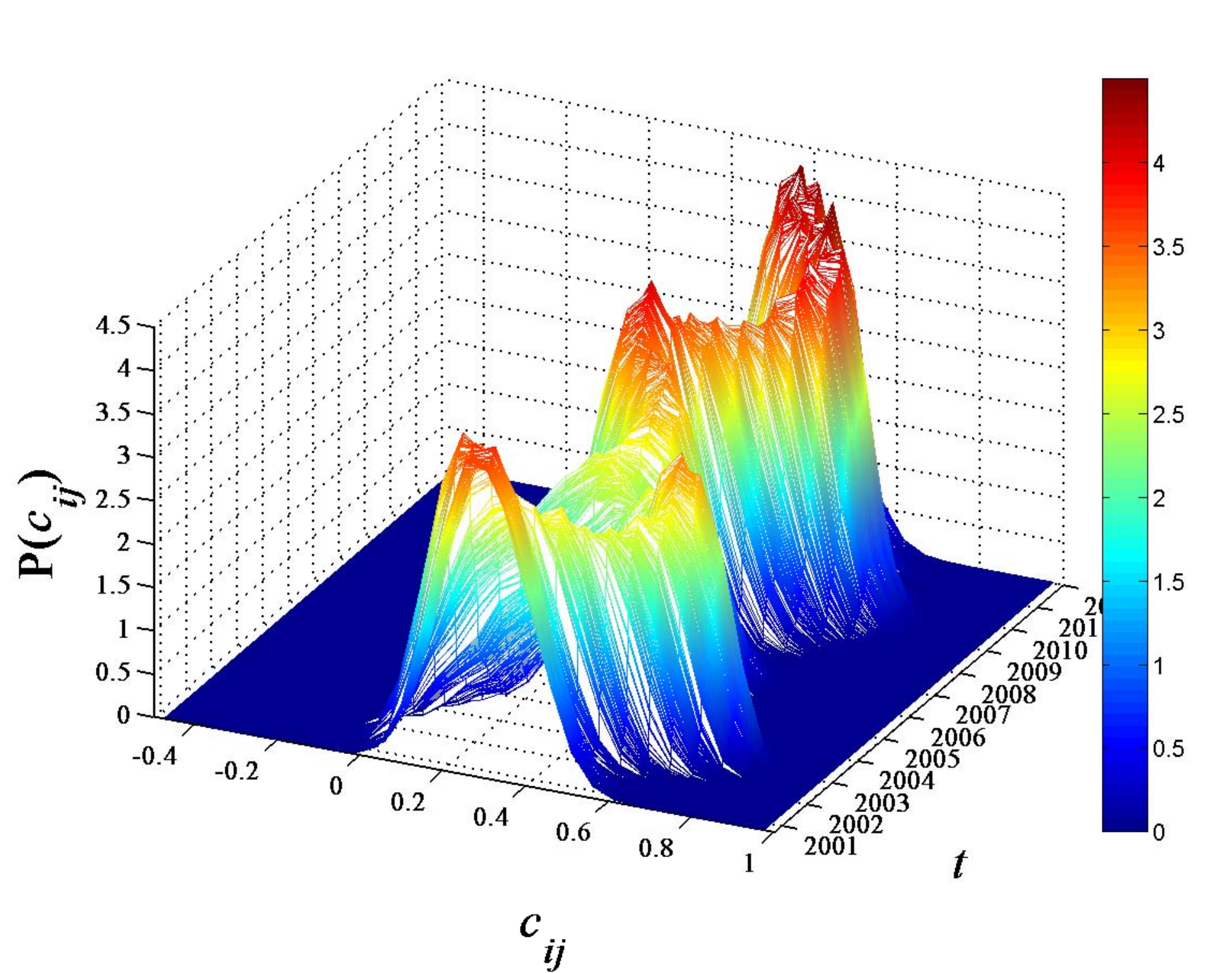}}\hspace{5mm}
\subfigure[][]{\label{Fig:return:PDF:correff-b}\includegraphics[width=8cm]{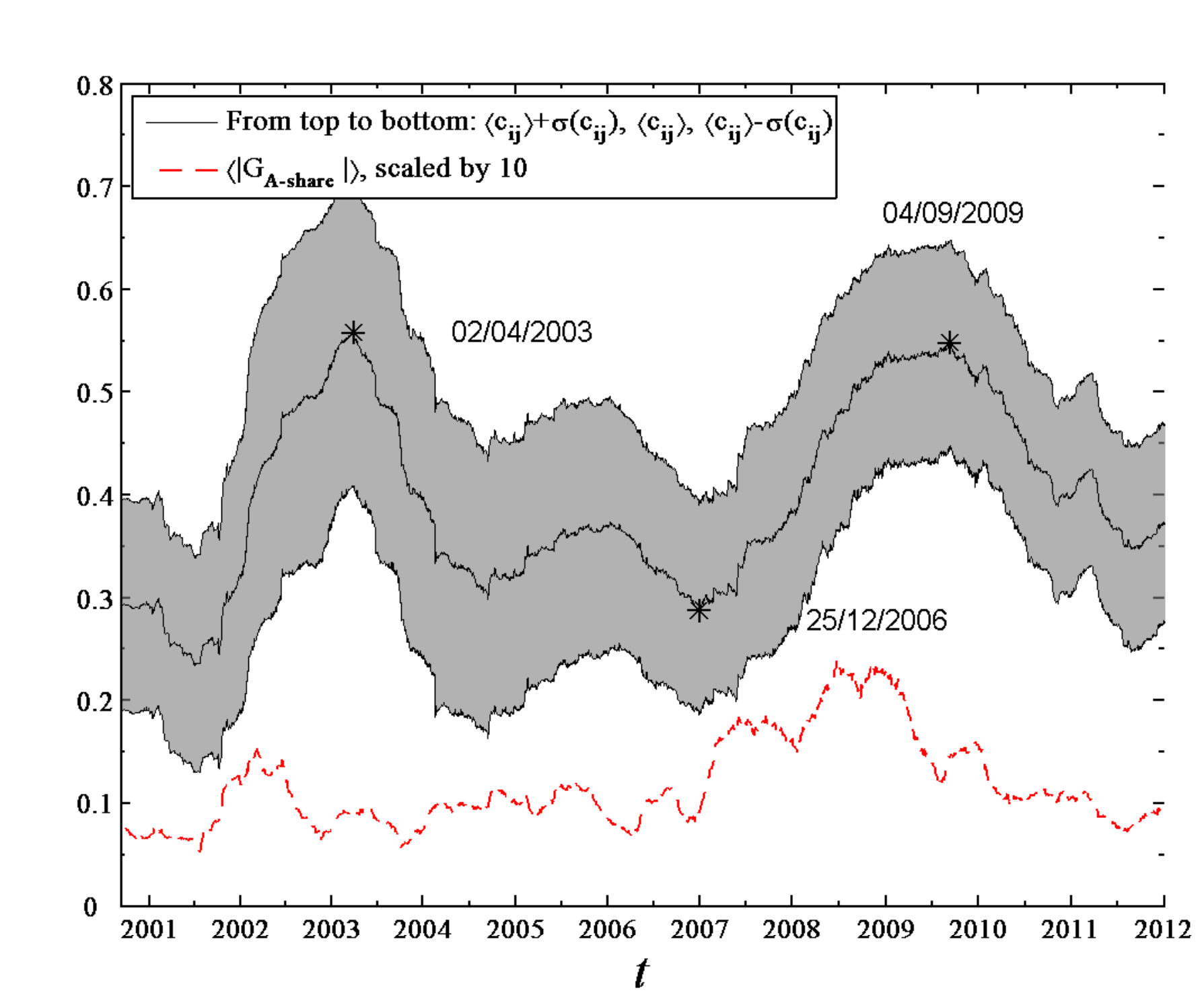}}
\caption{\subref{Fig:return:PDF:correff-a} Probability density function (PDF) $P(c_{ij})$ of the correlation coefficients calculated from the return series of 367 A-Share stocks evolved with the time $t$. \subref{Fig:return:PDF:correff-b} Mean correlation coefficient $\langle c_{ij} \rangle$ and mean absolute return $\langle |G_{A-share}| \rangle$ of the A-share Index of Shanghai Stock Exchange evolved with the time $t$. The black solid line in the middle of the shadow shows $\langle c_{ij} \rangle$ calculated from the return series of 367 A-Share stocks within a sliding window of length 400 days, and solid lines at the top and bottom of the shadow are $\langle c_{ij} \rangle + \sigma(c_{ij})$ and $\langle c_{ij} \rangle - \sigma(c_{ij})$, where $\sigma(c_{ij})$ is the standard deviation of the correlation coefficients. The red dashed line shows $\langle |G_{A-share}| \rangle$ of the A-share Index of Shanghai Stock Exchange calculated from the daily records of the A-share Index within a sliding window of length 100 days, scaled by a factor 10.}
\label{Fig:return:PDF:correff}
\end{figure}

We first analyze the distribution of the elements $c_{ij}$ of the correlation matrix to capture the statistical properties of the correlation coefficients. In Fig.~\ref{Fig:return:PDF:correff-a}, the probability density function (PDF) $P(c_{ij})$ of the correlation coefficients evolved with time $t$ is shown. We observe that the center of the distribution clearly deviates from zero for the whole range of $t$. The values of the coefficient $c_{ij}$, at which the peaks of $P(c_{ij})$ are located, are significantly positive and vary with the time $t$. The peaks of $P(c_{ij})$ show two local maxima of $c_{ij}$ as $t$ approaches 2003 and 2009, and appear at relatively small $c_{ij}$ for other $t$.

The Chinese stock market suffered a big crash after the release of the policy of state-held shares sale in listed companies in 2001, and the collapse of the internet bubble also took place in 2000-2001. In 2008, the global financial crisis burst out, and hit the stock markets around the world, certainly including the Chinese stock market. Considering that the length of the moving window is about two years, the correlations between the stock returns are significantly increased in the time windows 2001-2003 and 2008-2009 (for the specific dates please see Subsection 3.2). This indicates that stock price variations are more likely to be correlated around the market crashes.

\subsection{Mean correlation coefficient}

To further verify the dependence of the stock correlations on the time $t$, we compute the mean correlation coefficient $\langle c_{ij} \rangle$ in the moving window. Figure~\ref{Fig:return:PDF:correff-b} plots $\langle c_{ij} \rangle$ as a function of the evolving time $t$, and it strongly fluctuates during the whole range of $t$. According to the shape of $\langle c_{ij} \rangle$ shown in the figure, it exhibits two local maxima on 02/04/2003 and 04/09/2009 and a local minimum on 25/12/2006. The moving windows correspond to the two maxima are from 30/07/2001 to 02/04/2003 and from 17/01/2008 to 04/09/2009, and for the minimum is from 10/05/2005 to 25/12/2006. The date 30/07/2001 was close to the date 26/07/2001 on which the policy of state-held shares sale was formally implemented, and 17/01/2008 was near the date 21/01/2008 on which the Shanghai Stock Exchange Index dropped more than 5\% followed by a decline over 7\% the next day.

The volatility of the A-share Index of Shanghai Stock Exchange, quantified as the mean absolute returns within the moving window of 100 days length, is also illustrated in Fig.~\ref{Fig:return:PDF:correff-b}. In the periods from 30/07/2001 to 02/04/2003 and from 17/01/2008 to 04/09/2009 the stock market was strongly fluctuating, while in the period from 10/05/2005 to 25/12/2006 the market was in a relatively calm state. In comparison with the variation of $\langle c_{ij} \rangle$, one may conclude that stock correlations are more prominent in volatile periods, showing larger values of $\langle c_{ij} \rangle$ than those in calm periods.

\section{Dynamic behaviors of eigenvalues and their explanations of system variance}

\subsection{Distribution of eigenvalues}

\begin{figure}
\centering
\subfigure[][]{\label{Fig:return:PDF:correig-a}\includegraphics[width=7.5cm]{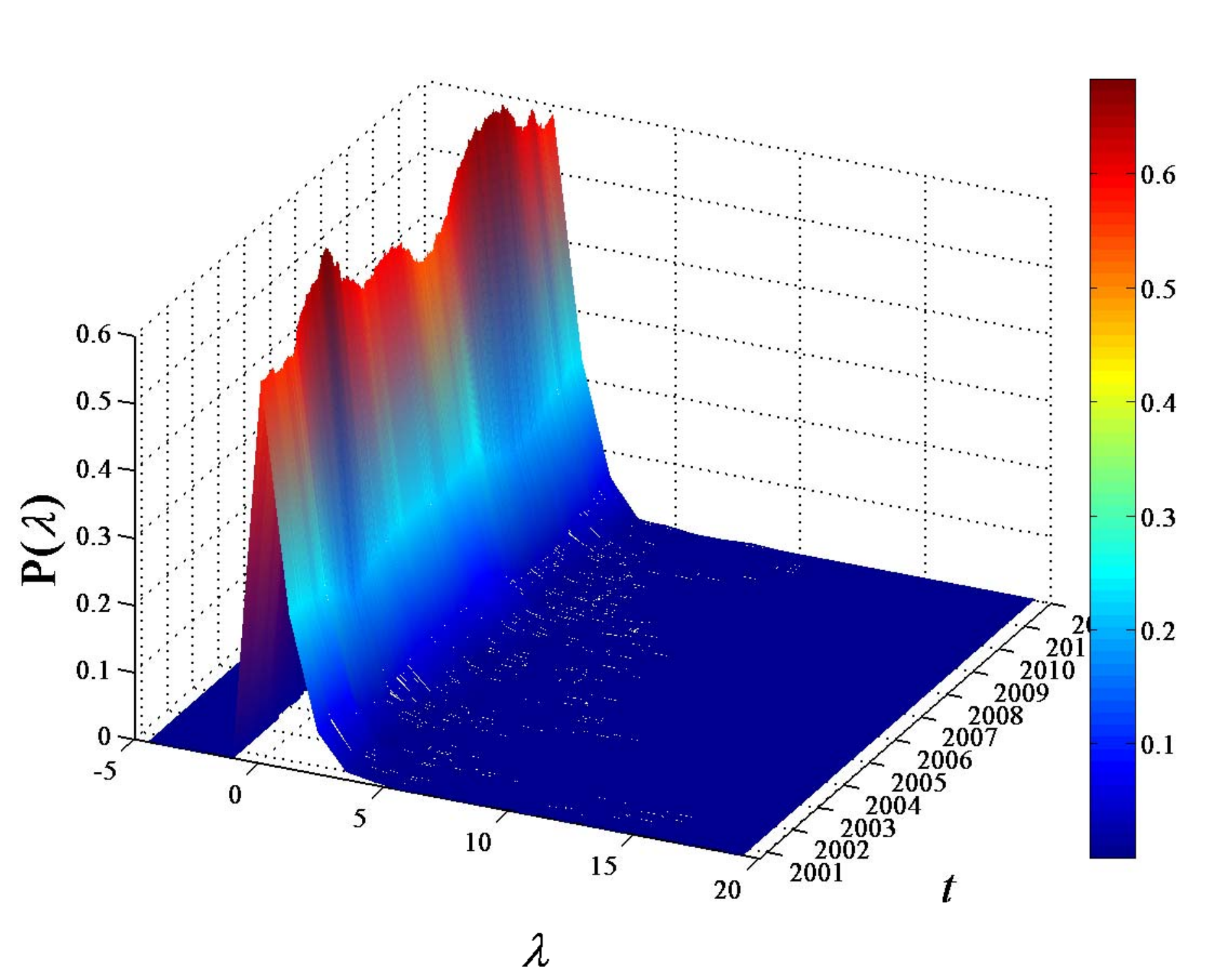}}\hspace{5mm}
\subfigure[][]{\label{Fig:return:PDF:correig-b}\includegraphics[width=7cm]{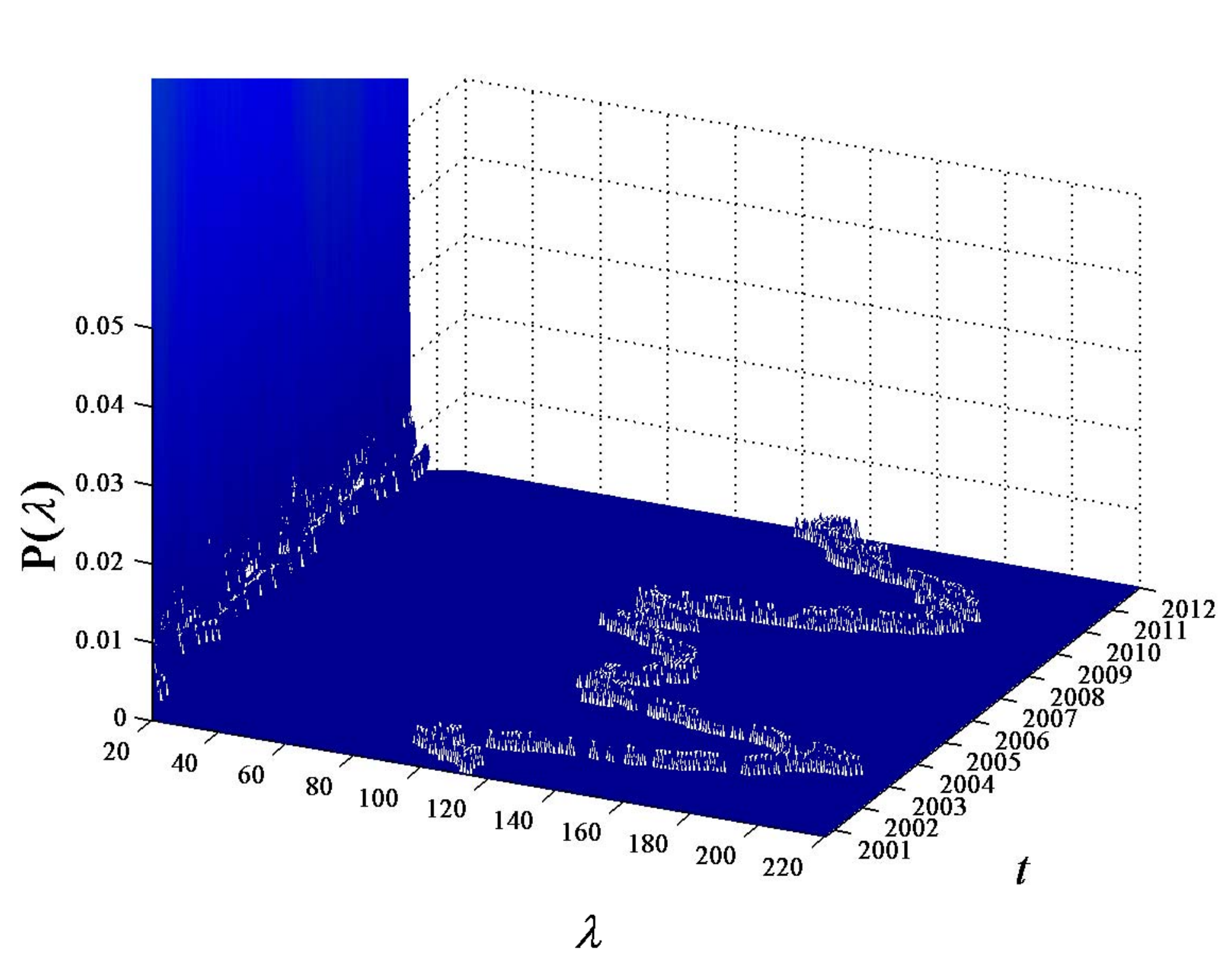}}
\caption{\subref{Fig:return:PDF:correig-a} Probability density function (PDF) $P(\lambda)$ of the eigenvalues obtained from the correlation matrix of the return series of 367 A-Share stocks evolved with the time $t$. \subref{Fig:return:PDF:correig-b} Partial enlarged drawing of figure~\ref{Fig:return:PDF:correig-a} for $\lambda >20$.}
\label{Fig:return:PDF:correig}
\end{figure}

We compute the eigenvalues of the correlation matrix $C$ with $N\times N$ elements, and denote them as $\lambda_k$, $k=1,\cdots,N$, and $\lambda_1 > \lambda_2 > \cdots > \lambda_N$. We investigate the probability density function (PDF) of the eigenvalues and its variation over time $t$. In Fig.~\ref{Fig:return:PDF:correig-a}, the PDF $P(\lambda)$ for $\lambda \leq 20$ evolved with $t$ is plotted. The peaks of $P(\lambda)$ show larger values for $t$ around 2003 and 2009 than those for other $t$. The $P(\lambda)$ for large eigenvalues $\lambda>20$ is plotted in Fig.~\ref{Fig:return:PDF:correig-b}. The largest eigenvalue evolves with time $t$, and shows larger values, i.e., $\lambda_1 > 200$, in the time windows 2001-2003 and 2008-2009. This phenomenon consists with the unveiling of two local maxima of $<c_{ij}>$ in the moving windows from 30/07/2001 to 02/04/2003 and from 17/01/2008 to 04/09/2009.

\subsection{Number of eigenvalues that significantly deviate from random correlation matrix}

\begin{figure}[htb]
\centering
\subfigure[][]{\label{Fig:return:correig:shuffle-a}\includegraphics[width=8cm]{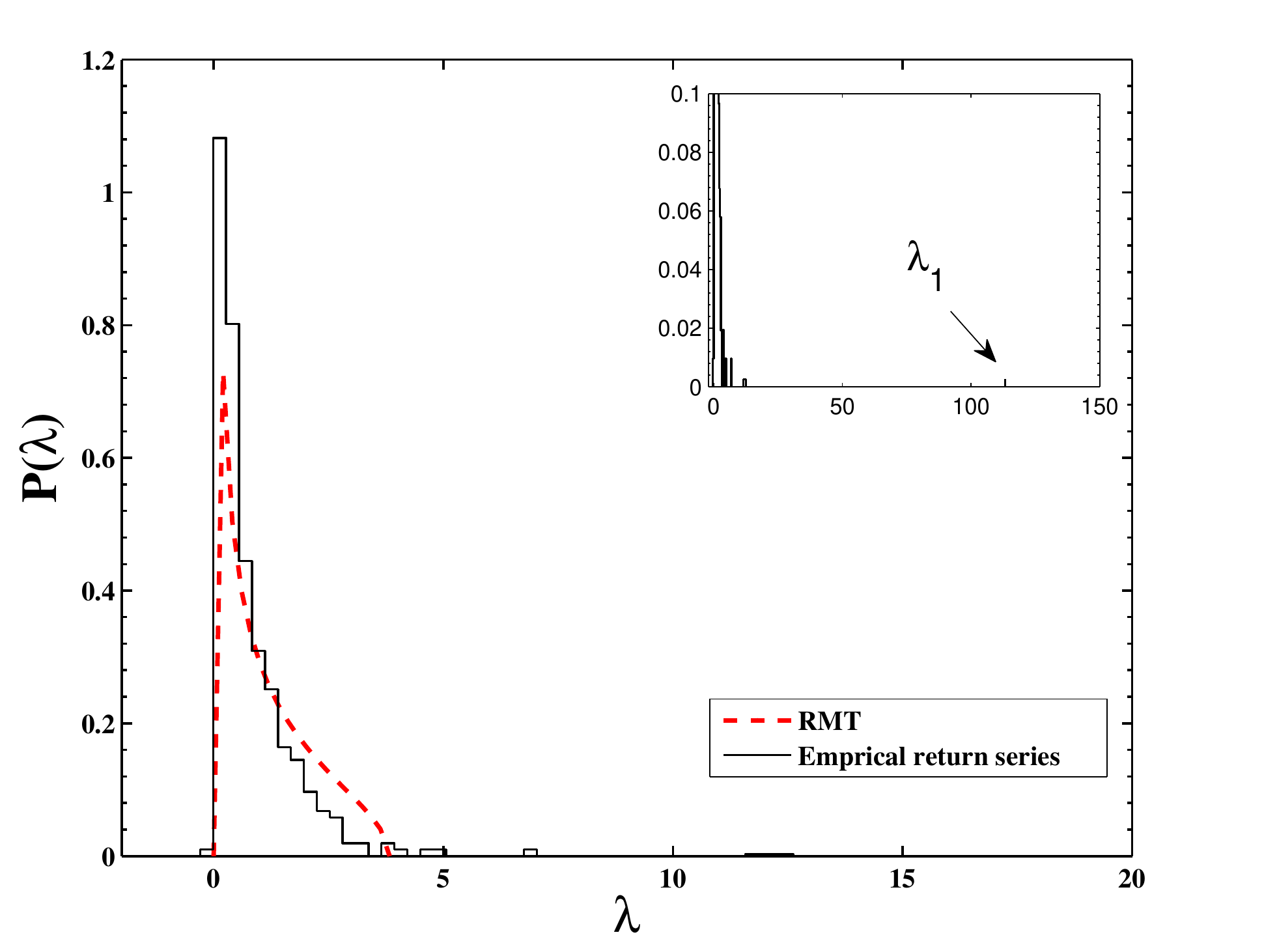}}\hspace{5mm}
\subfigure[][]{\label{Fig:return:correig:shuffle-b}\includegraphics[width=7cm]{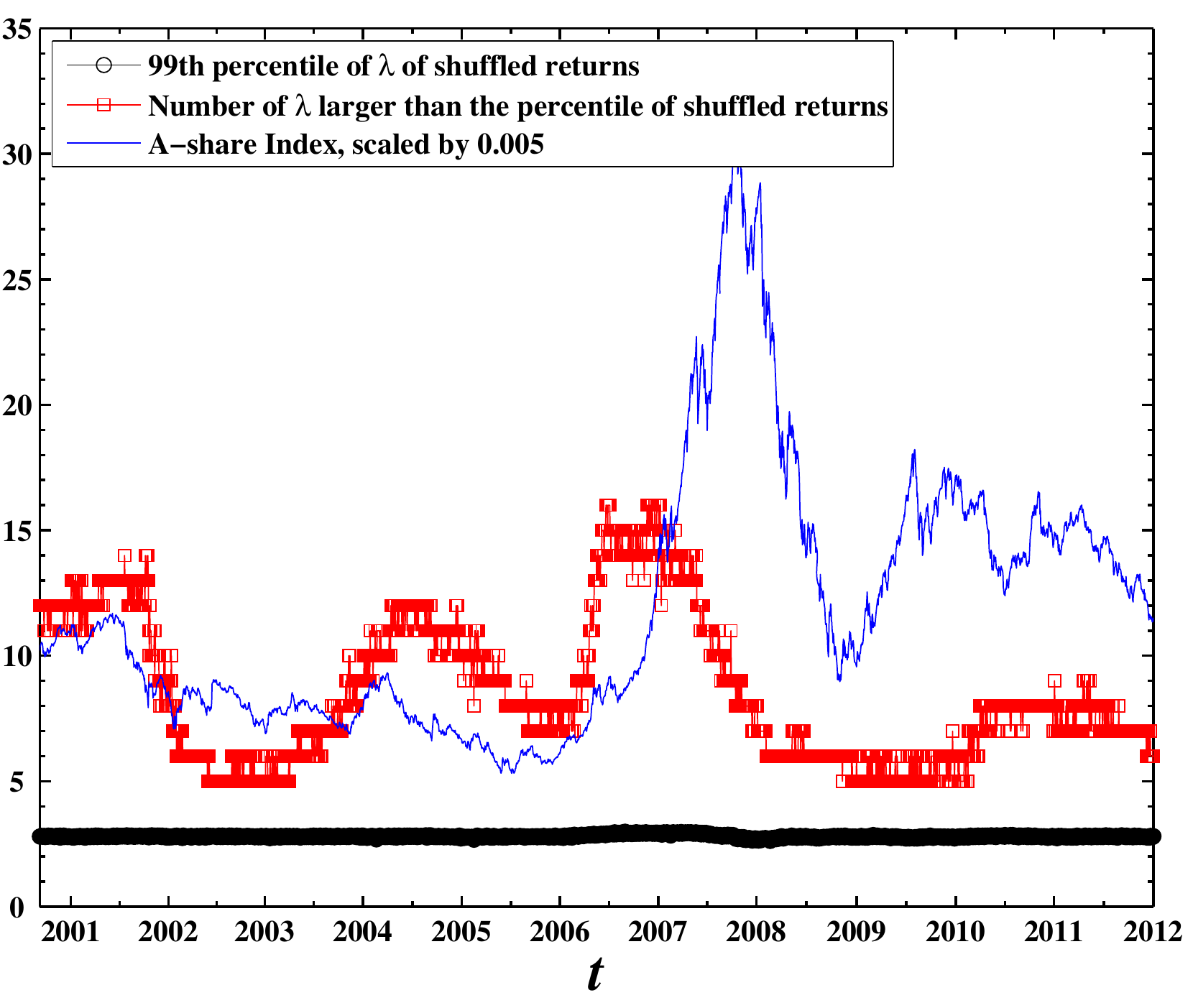}}
\caption{\subref{Fig:return:correig:shuffle-a} Probability density function (PDF) $P(\lambda)$ of the eigenvalues of the correlation matrix constructed from the return series of 367 A-Share stocks in the first moving window form 04/01/1999 to 06/09/2000. The dotted line is the RMT result obtained from Eq.~\ref{equ_RMT}. The inset shows the largest eigenvalue $\lambda_1$ of the empirical return series, which is much larger than the upper bound $\lambda_{max}$ of RMT. \subref{Fig:return:correig:shuffle-b} Comparison between the eigenvalues calculated from the empirical return series and the shuffled return series. The black circled line is the 99th percentile of the eigenvalues of the shuffled return series. The return series in each moving window is randomized by shuffling for 10 times. The red squared line is the number of the empirical eigenvalues significantly larger than those of the shuffled data, which are the eigenvalues larger than the 99th percentile of the eigenvalues of the shuffled data. The blue solid line shows the A-share Index of Shanghai Stock Exchange, scaled by a factor 0.005.}
\end{figure}

In the observation of $P(\lambda)$, we note that there exist large eigenvalues obviously large than the eigenvalues of the random correlation matrix. To compare the difference between the eigenvalues of the stock correlation matrix and those of the random correlation matrix, we show the analytical result of the random matrices following Ref.\cite{Sengupa-Mitra-1999-PRE}. For the correlation matrix of $N$ random time series of length $L$, the PDF $P(\lambda)$ of the eigenvalues $\lambda$ in the limit $N\rightarrow\infty$ and $L\rightarrow\infty$ is given by
\begin{equation}
P(\lambda)=\frac{Q}{2\pi}\frac{\sqrt{(\lambda_{max}-\lambda)(\lambda-\lambda_{min})}}{\lambda}, \label{equ_RMT}
\end{equation}
where $Q\equiv L/N >1$, and $\lambda$ is within the bounds $\lambda_{min}\leq\lambda\leq\lambda_{max}$. $\lambda_{min}$ and $\lambda_{max}$ are the minimum and maximum eigenvalues of the random correlation matrix, which are given by
\begin{equation}
\lambda_{min,max}=1+\frac{1}{Q}\mp2\sqrt{\frac{1}{Q}}.
\label{equ_RMT_eigenvalue}
\end{equation}
In Fig.~\ref{Fig:return:correig:shuffle-a}, we plot $P(\lambda)$ of the random correlation matrix with finite $L=400$ and $N=367$, the same as those of the stock return series. Within the bounds $[\lambda_{min},\lambda_{max}]$, $P(\lambda)$ of the correlation matrix constructed from the empirical return series in the first moving window (black solid line) is consistent with the analytical result of Eq.~\ref{equ_RMT} (red dashed line). There also exist some deviations of large eigenvalues. In particular, the largest eigenvalue $\lambda_1\approx120$ shown in the inset of Fig.~\ref{Fig:return:correig:shuffle-a}, which is about 31 times larger than $\lambda_{max}=3.83$.

We next identify the eigenvalues of the stock correlation matrix which deviate from those of the random correlation matrix, and investigate their variations over time $t$. The analytical result of RMT is strictly valid for $N\rightarrow\infty$ and $L\rightarrow\infty$. Instead, we compare $\lambda$ of the stock correlation matrix with $\lambda$ of the correlation matrix constructed from $N=367$ uncorrelated time series with length $L=400$. The uncorrelated time series is generated by shuffling the empirical return series, in which the equal-time correlations between the original return series are destroyed. We compute the cross-correlations between these shuffled return series, and use this surrogate correlation matrix as a random correlation matrix. In Fig.~\ref{Fig:return:correig:shuffle-b}, black circled line denotes the 99th percentile of the eigenvalues calculated from the random correlation matrix. It stays relatively constant about 3 as the time $t$ evolves. This means that 99 percent of the eigenvalues of the random correlation matrix are less than this value.

If an eigenvalue of the empirical correlation matrix is larger than the 99th percentile of the eigenvalues generated from the shuffled return series, it is considered to be significantly larger than the eigenvalues of the random correlation matrix. In Fig.~\ref{Fig:return:correig:shuffle-b}, the number of the eigenvalues significantly larger than those of the random correlation matrix is plotted by the red square line. The number of empirical $\lambda$ significantly larger than $\lambda$ of random correlation matrix fluctuates over time $t$. For $t$ around two date points 02/04/2003 and 04/09/2009, it shows a minimal value about 5, while for $t$ around 25/12/2006, it shows a maximal value about 16. This means that the number of significant eigenvalues in the volatile periods close to 30/07/2001-02/04/2003 and 17/01/2008-04/09/2009 is lager than that in the calm period close to 10/05/2005-25/12/2006. To further illustrate the volatile and calm periods of the A-share market, we also plot the index composed of all A-share stocks in the figure. The crashes of 2001-2003 and 2008-2009 seem to start from middle 2001 and early 2008 respectively, and the following indices keep dropping for long periods of time. Between these two crashes, there exists a calm period from middle 2005 to late 2006, in which the A-share index shows a local minimum in middle 2005 and relatively small values till late 2006.

We give a cursory explanation for the above phenomenon. It can be easily proved that the sum of the eigenvalues of the stock correlation matrix is fixed to be the number of sample stocks, i.e, $\sum_{k=1}^{N} \lambda _k=N$. As shown in the distribution of the eigenvalues, the major portion of eigenvalues are distributed in the region $\lambda < 3$, and the large eigenvalues $\lambda >20$ close to the market crashes of 2001-2003 and 2008-2009 are prominently larger than those during the calm period. Therefore, the number of eigenvalues in-between $3<\lambda<20$ during crashes is less than calm periods. This may indicate that a few of the eigenvalues contain the information about the stock correlations when the market strongly fluctuates.

\subsection{Portion of system variance explained by eigenvalues}

The commonality among the stock returns can also be detected by the PCA method, which has a close link to the RMT method. In fact, the systemic risk measured by the collective behavior of the stock price movements based on PCA has been analyzed in many studies \cite{Zheng-Podobnik-Feng-Li-2012-SR,Billio-Getmansky-Lo-Pelizzon-2012-JFE,Kritzman-Li-Page-Rigobon-2011-JPM}. The PCA method decomposes the returns of a sample of stocks into several orthogonal principal components. The principal components $\zeta_k$ are uncorrelated, and satisfy the condition $<\zeta_k \zeta_l>=\lambda_k$ if $k=l$, where $\lambda_k$ is the $k$-th eigenvalue of the correlation matrix $C$ of stock returns. The standardized return of stock $i$, defined as $z_i=(G_i(t)- \langle G_i(t)\rangle)/\sigma_i$, can be expressed as a linear combination of the principal components $\zeta_k$
\begin{equation}
z_i=\sum_{k=1}^{N} L_{ik}\zeta_k, \label{equ_PCA_Com}
\end{equation}
where $N=367$ is the total number of stocks analyzed, and $L_{ik}$ is the component of $k$-th eigenvector corresponding to stock $i$, which is also known as the factor loading of $\zeta_k$ for stock $i$. The total variance of the return series is
\begin{equation}
\sigma^2=\sum_{i=1}^N \sum_{j=1}^N \sum_{k=1}^N \sigma_i \sigma_j L_{ik} L_{jk} \lambda_k , \label{equ_PCA_Var}
\end{equation}
in which the total variance is decomposed into the orthogonal factor loadings $L$ and the eigenvalues $\lambda$. For the periods that stocks are highly correlated and connectively volatile, a small number $n<N$ of eigenvalues can explain most of the volatility in the system.

The cumulative risk fraction (CRF) is generally used to quantify the proportion of total variance explained by the first $n$ principal components \cite{Billio-Getmansky-Lo-Pelizzon-2012-JFE}, also known as absorption ratio in ~\cite{Kritzman-Li-Page-Rigobon-2011-JPM}. It is defined as
\begin{equation}
h_n=\frac{\sum_{k=1}^n \lambda_{k}}{\sum_{k=1}^N \lambda_{k}}, \label{equ_CRF}
\end{equation}
where $\lambda_{k}$ is the $k$-th eigenvalue, $\lambda_1> \lambda_2 >\cdots >\lambda_N$. Since the PCA is done through the decomposition of the correlation (covariance) matrix of return (standardized return) series, the total variance of the system explained by all $N$ principal components is quantified as $\sum_{k=1}^{N} \lambda_{k}$. The variance associated with the first $n$ principal components is quantified as $\sum_{k=1}^{n} \lambda_{k}$. The CRF is the ratio of these two quantities.

\begin{figure}[htb]
\centering
\includegraphics[width=8cm]{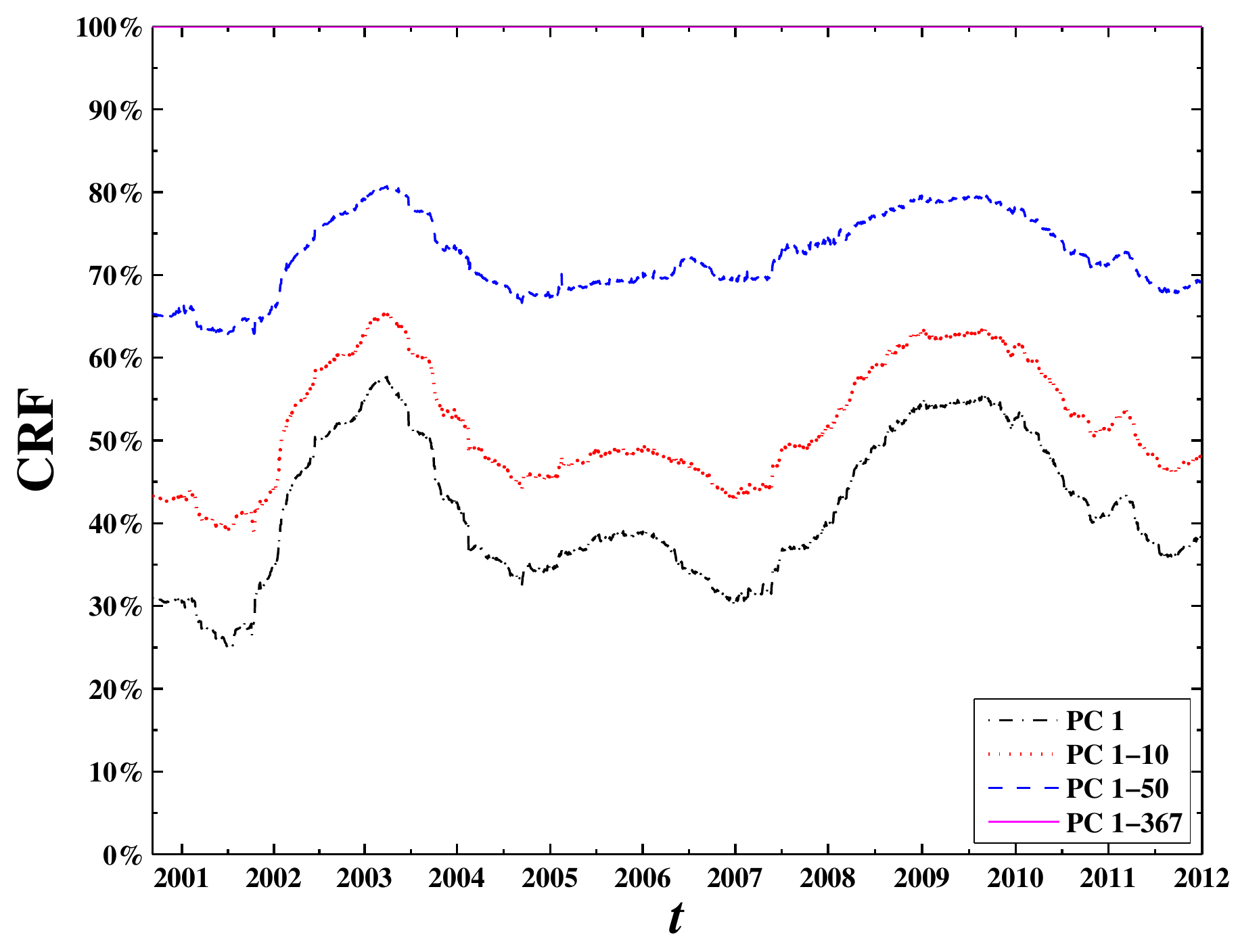}\hspace{5mm}
\caption{\label{Fig:return:CRF} Cumulative Risk Fraction (CRF) measured by the eigenvalues obtained from the correlation matrix of the return series of 367 A-Share stocks evolved with the time $t$. Different lines correspond to the proportions of total variance explained by PC 1, PC 1-10, PC 1-50, and PC 1-367. PC 1 denotes the principal component corresponding to the largest eigenvalue $\lambda_1$.}
\end{figure}

In Fig.~\ref{Fig:return:CRF}, the CRFs for $n=1, 10, 50, 367$ are shown as a function of the evolving time $t$. The CRF displays two local maxima at $t$ nearby 02/04/2003 and 04/09/2009, at which it can explain more than 50\%, 60\%, and 80\% of the total variance for $n=$1, 10, and 50 respectively. This indicates that the stocks are highly correlated in the moving windows  from 30/07/2001 to 02/04/2003 and from 17/01/2008 to 04/09/2009, in which the majority of the stock returns tend to move together. Thus the stock market is at a high level of systemic risk. We also observe that the CRF displays a relatively small value in the moving window from 10/05/2005 to 25/12/2006, in which the stocks are less correlated. These results are coincident with those observed in the mean correlation coefficient.

\section{Evolutions of statistical properties of eigenvectors and their interpretations}

\subsection{Evolution of the eigenvector components grouped in conventional industries}

To analyze the information contained in the deviating eigenvectors, we first investigate the contributions of the eigenvector components grouped in conventional industries. According to the China Securities Regulatory Commission (CSRC) industry code, the stocks traded on Shanghai Stock Exchange are grouped into A-M conventional industries. Table~\ref{TB:Industry:Summary} presents summary statistics of the 22 industries, including the industry codes, industry names, and the number of chosen stocks belonging to each industry. For each deviating eigenvector $u^k$, with element $u_i^k$ as the component of the $k$-th eigenvector corresponding to stock $i$, we calculate the contribution of each industry group
\begin{equation}
X_{l}^{k}=\sum_{i=1}^{N_l} \frac{1}{N_l}(u_i^k)^{2}, \label{equ_CI}
\end{equation}
where $N_l$ is the number of stocks belonging to industry group $l$, $l=1,\cdots,22$. The measure of $X_{l}^{k}$ is analogous to the analysis of wave function in disordered systems, and firstly introduced to financial data analysis in Ref.~\cite{Gopikrishnan-Rosenow-Plerou-Stanley-2001-PRE}.

\begin{table}[htp]
 \centering
 \caption{A-M conventional industries grouped based on the China Securities Regulatory Commission (CSRC) industry code. The basic information includes the industry code, full name of the industry, and the number of chosen stocks belonging to each industry.} \label{TB:Industry:Summary}
\begin{tabular}{lp{4cm}c}
  \hline
  Industry code & Industry & Number of stocks \\
  \hline
    A  & Agriculture           & 4  \\
    B  & Mining                & 7  \\
    C0 & Food \& beverage      & 21 \\
    C1 & Textiles \& apparel   & 12 \\
    C2 & Timber \& furnishings & 0  \\
    C3 & Paper \& printing     & 5  \\
    C4 & Petrochemicals        & 31 \\
    C5 & Electronics           & 10 \\
    C6 & Metals \& non-metals  & 26 \\
    C7 & Machinery             & 52 \\
    C8 & Pharmaceuticals       & 19 \\
    C99& Other manufacturing                & 2 \\
    D  & Utilities                          & 19 \\
    E  & Construction                       & 6 \\
    F  & Transportation                     & 14 \\
    G  & IT                                 & 18 \\
    H  & Wholesale \& retail trade         & 54 \\
    I  & Finance \& insurance              & 2 \\
    J  & Real estate                        & 43 \\
    K  & Social services                    & 11 \\
    L  & Communication \& cultural industry& 4 \\
    M  & Comprehensive                      & 7 \\
  \hline
\end{tabular}
\end{table}

\begin{figure}
\centering
\subfigure[][]{\label{Fig:return:eigenvector-a}\includegraphics[width=4cm]{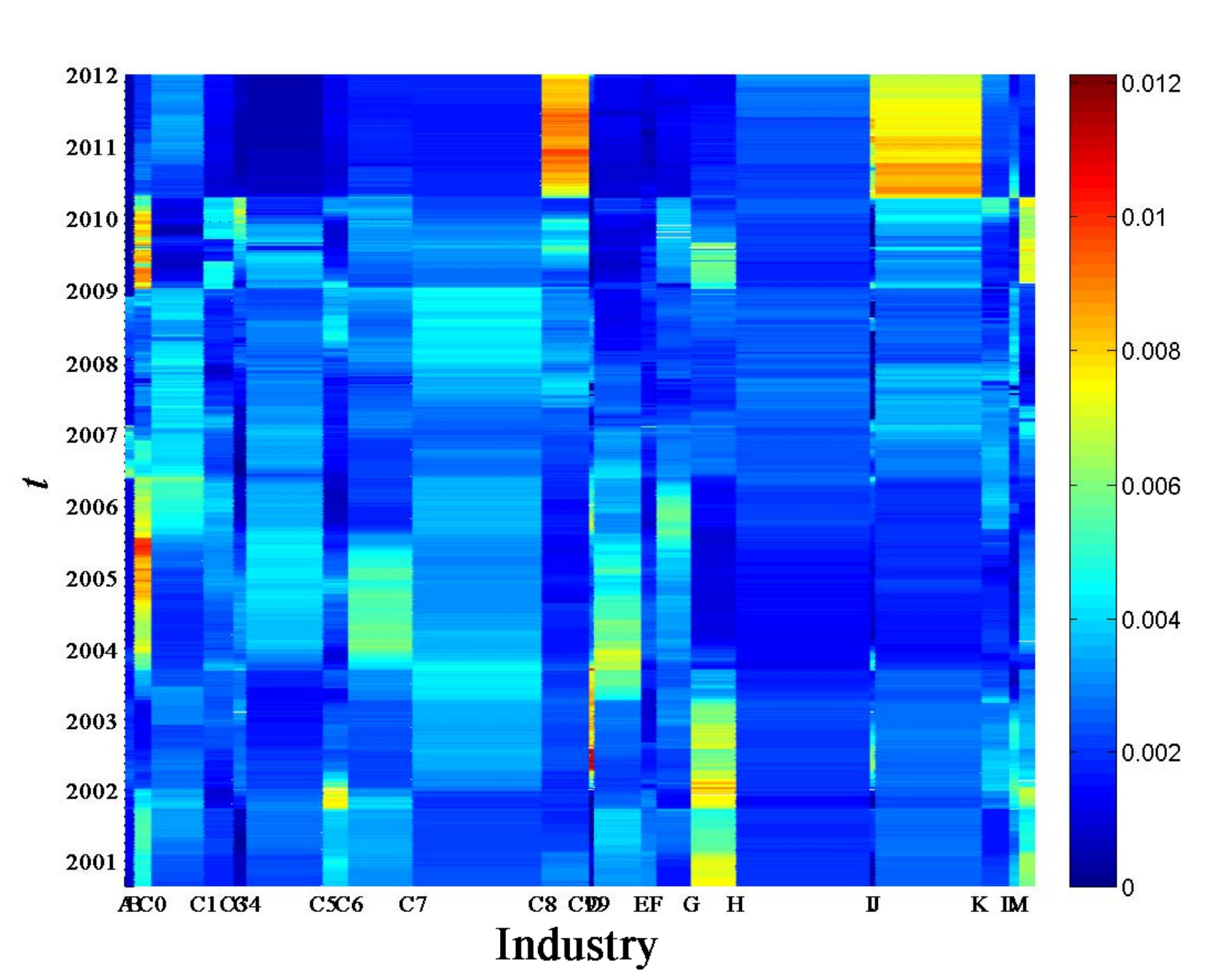}}\hspace{5mm}
\subfigure[][]{\label{Fig:return:eigenvector-b}\includegraphics[width=4cm]{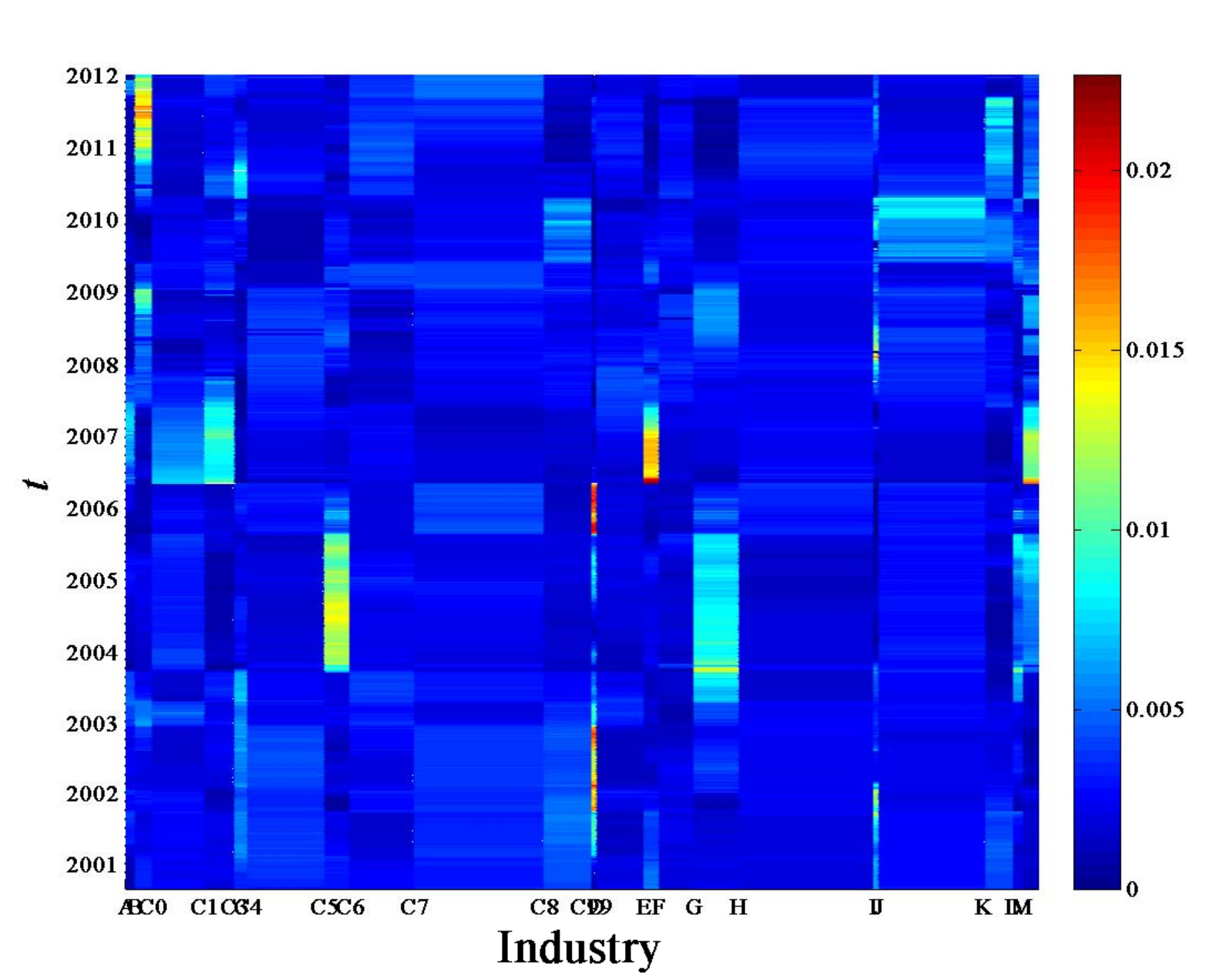}}
\subfigure[][]{\label{Fig:return:eigenvector-c}\includegraphics[width=4cm]{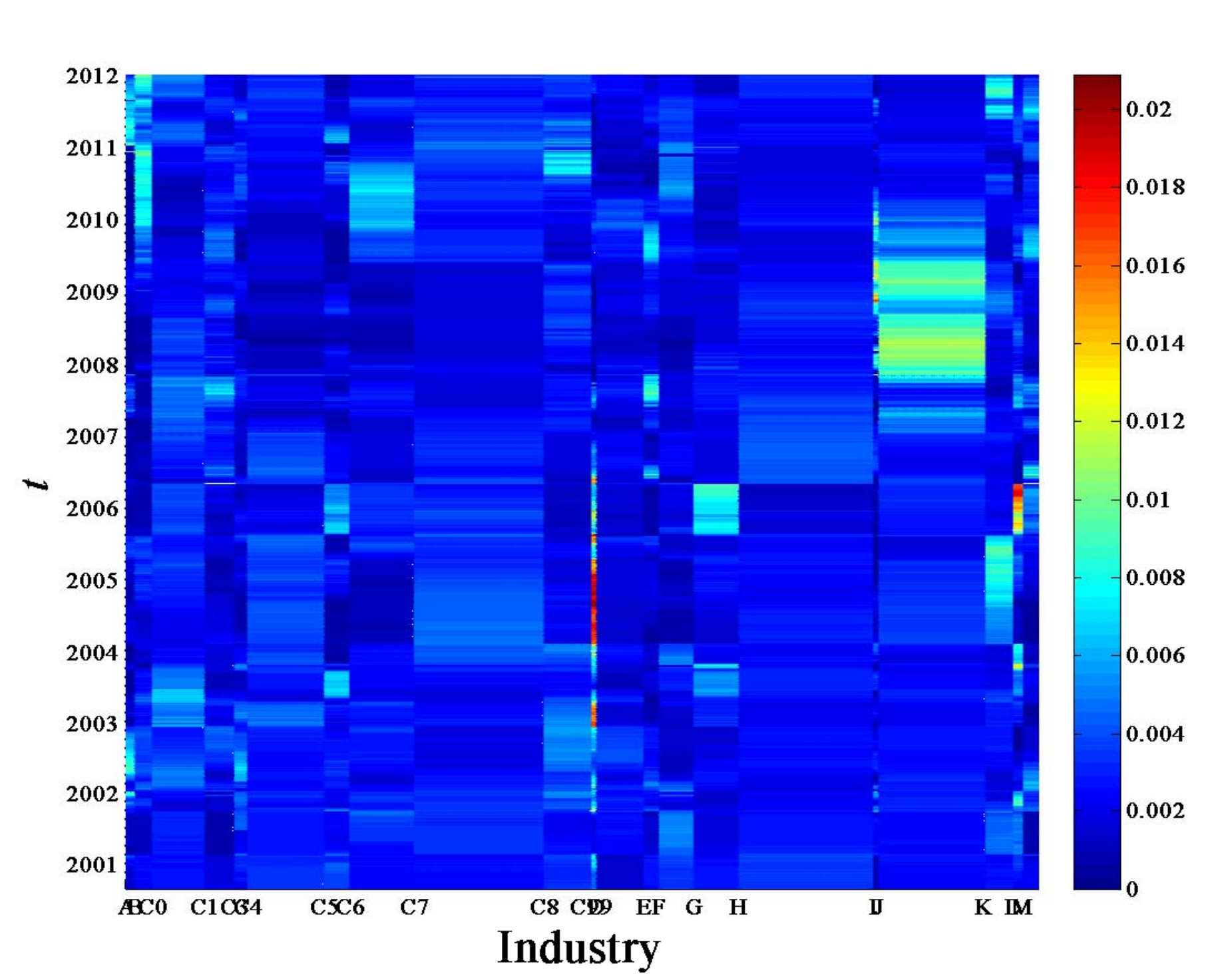}}\hspace{5mm}
\subfigure[][]{\label{Fig:return:eigenvector-d}\includegraphics[width=4cm]{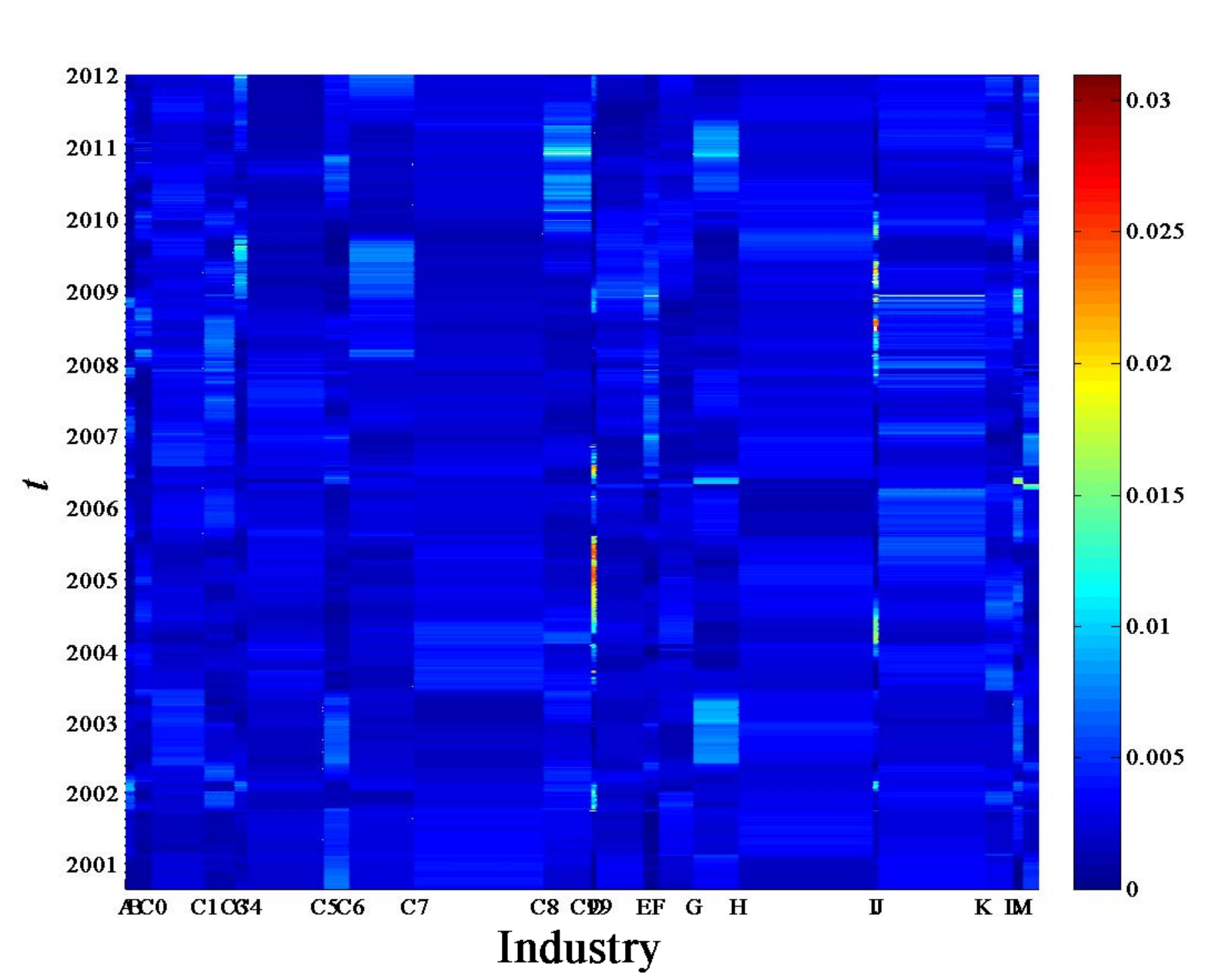}}
\caption{Contribution $X_{l}^{k}$ of conventional industries for \subref{Fig:return:eigenvector-a} $u^2$, \subref{Fig:return:eigenvector-b} $u^3$, \subref{Fig:return:eigenvector-c} $u^4$, and \subref{Fig:return:eigenvector-d} $u^5$ obtained from the correlation matrix of the return series of 367 A-Share stocks evolved with time $t$.}
\label{Fig:return:eigenvector}
\end{figure}

We find that $X_{l}^{k}$ for the largest eigenvector $u^1$ universally show large values among different industries, which means that almost all the industries have significant contributions to $u^1$. It is quite robust for different $t$. Fig.~\ref{Fig:return:eigenvector} shows $X_{l}^{k}$ for other deviating eigenvectors $u^2$, $u^2$, $u^4$ and $u^5$ evolved with time $t$. The participants of the eigenvectors listed in the horizontal axis are 367 stocks. The stocks belong to industry group $l$ are endowed with the same value of $X_{l}^{k}$, and ranked by their capitalizations on the ending date of the sample data.

$X_{l}^{k}$ shows different patterns in the periods divided by the date points 02/04/2003, 25/12/2006, and 04/09/2009. In addition, $X_{l}^{k}$ before and after two date points 13/01/2009 and 11/05/2010, which are the ending dates of the moving windows started from 30/05/2007 and 16/09/2008 respectively, show remarkably different patterns. These discrete patterns can be easily observed for $u^2$ and $u^3$. The Shanghai Stock Exchange fell 6.5\% on 30/05/2007, which was caused by an increase in the stamp tax on stock transactions to 0.3\% from 0.1\%. The bankruptcy of Lehman Brothers on 14/09/2008 indicated that the financial crisis entered an acute phase, and the Chinese stock market started to be affected by the global financial crisis after that, showing a 4.5\% fall on 16/09/2008. Therefore, we choose the ending dates of these two moving windows, i.e., 13/01/2009 and 11/05/2010, as additional dividing dates. The date points 02/04/2003, 25/12/2006, 13/01/2009, and 11/05/2010 are picked as coarse-grained dividing points.

We next analyze the contributions of industries in different time periods separated by the four dividing dates. As shown in Fig.~\ref{Fig:return:eigenvector}, $u^2$ and $u^5$ shows large values of $X_{l}^{k}$ for the electronics and IT industries respectively in the first period from 06/09/2000 to 02/04/2003. In the following period from 02/04/2003 to 25/12/2006, mining, electronics, and real estate industries have significant contributions to $u^2$, $u^3$, and $u^5$ respectively. Real estate industry is a significant contributor of $u^4$ in the periods from 25/12/2006 to 13/01/2009, and of $u^3$ and $u^4$ from 13/01/2009 to 11/05/2010. In the last period from 11/05/2010 to 30/12/2011, both real estate and pharmaceuticals industries have significant contributions to $u^2$, and mining industry is a significant contributor of $u^3$. It is worth noting that $X_{l}^{k}$ of finance \& insurance and other manufacturing industries display large values for $u^2$-$u^5$. We neglect their contributions to the deviating eigenvectors, since there are only small numbers of chosen stocks belonging to these two industries.

\subsection{Interpretation of largest eigenvector $u^1$}

\begin{figure}[htb]
\centering
\includegraphics[width=8cm]{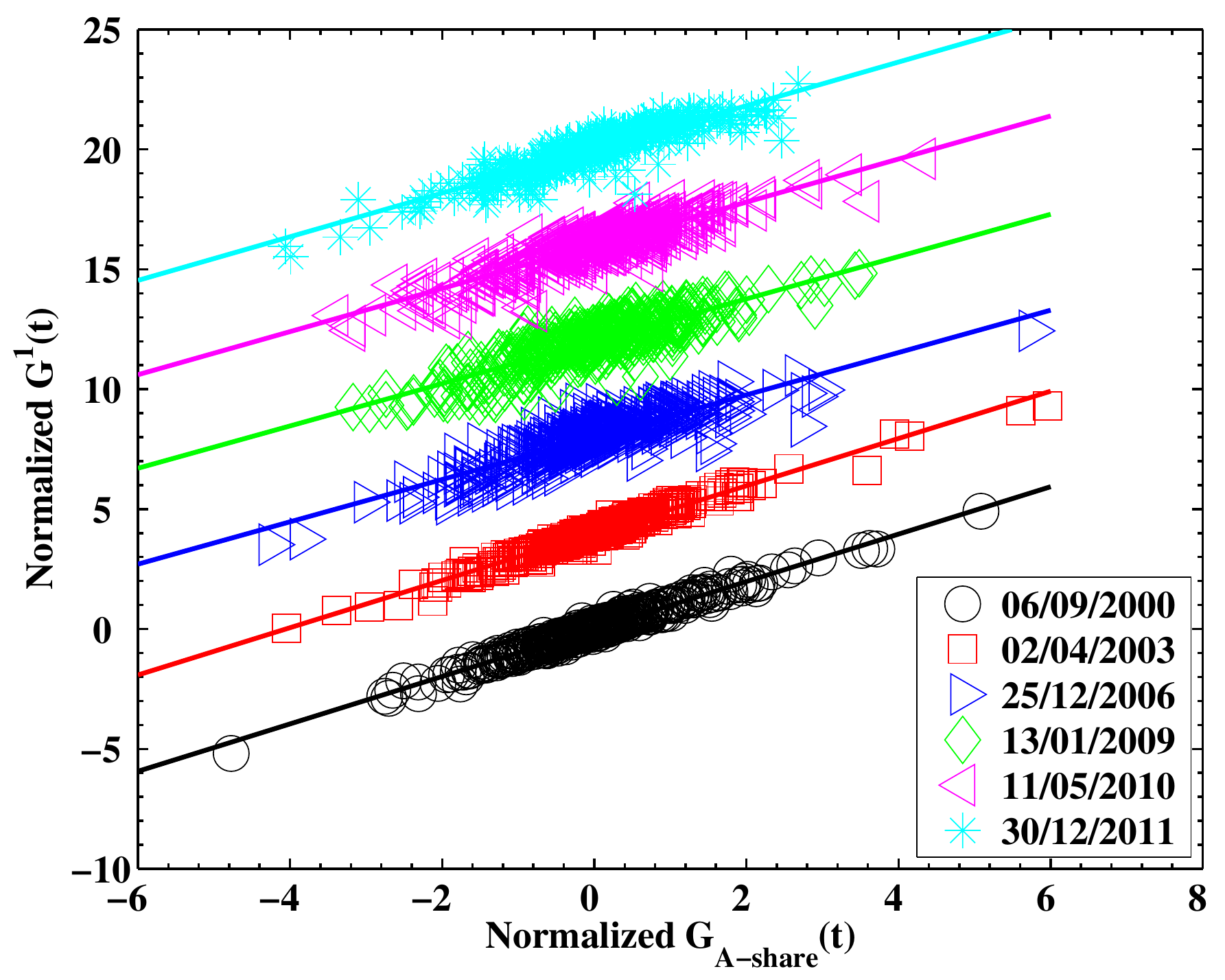}\hspace{5mm}
\caption{\label{Fig:return:Gmax-GshareA} Projection $G^1(t)$ of the 367 stock returns on the largest eigenvector $u^1$ obtained from the correlation matrices of the return series in the moving windows ended on 06/09/2000, 02/04/2003, 25/12/2006, 13/01/2009, 11/05/2010, and 30/12/2011, as a function of the return of the A-share Index of Shanghai Stock Exchange. The A-share Index is composed of all the A-share stocks listed on the Shanghai Stock Exchange. Curves are removed for clarity. A linear regression between the two normalized axes for different moving windows yields slopes: 0.99, 0.99, 0.88, 0.88, 0.90, and 0.91.}
\end{figure}

We calculate the projection of the stock returns $G_j(t)$ on the largest eigenvector $u^1$
\begin{equation}
G^1(t)=\sum_{j=1}^{N} u_j^1 G_j(t), \label{equ_projection}
\end{equation}
where $u_j^1$ is the component of $u^1$ corresponding to stock $j$, and $N$ is the number of sample stocks. In Fig.~\ref{Fig:return:Gmax-GshareA}, we plot $G^1(t)$ against the return of the A-share Index of Shanghai Stock Exchange $G_{A-share}(t)$ for the moving windows ended on 06/09/2000, 02/04/2003, 25/12/2006, 13/01/2009, 11/05/2010, and 30/12/2011. The A-share Index is composed of
all A-share stocks traded on Shanghai Stock Exchange. The projection $G^1(t)$ can be well fitted by a linear fit, which shows a narrow scatter around the fitted line in figure. The slope is about $0.93\pm0.06$, with a slight quantitative difference for different moving windows. The significant linear correlation between $G^1(t)$ and $G_{A-share}(t)$ indicates that the largest eigenvalue can be interpreted as quantifying market-wide influence on all stocks, and it remains quite robust to the variance of $t$. In fact, all the components of $u^1$ are positive in our study, and similar results are revealed in \cite{Shen-Zheng-2009a-EPL}. The A-share Index is a capitalization-weighted average of the prices of all A-share stocks, and large components of $u^1$ are universally distributed among all stocks in Fig.~\ref{Fig:return:eigenvector-a}. Thus it would be no surprise to observe the significant correlation between $G^1(t)$ and $G_{A-share}(t)$.

\subsection{Interpretation of deviating eigenvectors $u^2$-$u^5$}

We have offered an overall observation of the contributions of industry groups. For the interpretation of $u^2$-$u^5$, we further analyze the component stocks which significantly contribute to each deviating eigenvector in different time periods divided by the date points 02/04/2003, 25/12/2006, 13/01/2009, and 11/05/2010. The minor adjustments of the dividing dates centered around them do not significantly change the results. Table~\ref{TB:return-eigenvector-1}-\ref{TB:return-eigenvector-5} show the stocks and industry groups corresponding to the largest ten components of the deviating eigenvectors $u^2$, $u^3$, $u^4$, and $u^5$ by the average ranks of the eigenvector components in different time periods. We rank the components according to their eigenvector component values, and average the ranks of the components over the moving windows with ending dates from 06/09/2000 to 02/04/2003, from 02/04/2003 to 25/12/2006, from 25/12/2006 to 13/01/2009, from 13/01/2009 to 11/05/2010, and from 11/05/2010 to 30/12/2011. The components with the smallest ten average ranks are picked as the largest ten components. The largest ten components correspond to ten stocks which significantly contribute to the relevant eigenvectors.

If one looks carefully at the stock codes of the largest ten components, dynamic evolutions of conventional stock industries are remarkably observed. The stocks belonging to the industries which have significant contributions to distinct eigenvectors also appear in their largest ten components. For the moving windows with ending dates in the period from 06/09/2000 to 02/04/2003, as shown in Table ~\ref{TB:return-eigenvector-1}, among the largest ten components of $u^2$ five stocks belong to IT industry and one stock belongs to electronics industry, and for $u^3$ four stocks belong to machinery industry and two stocks belong to petrochemicals industry. In the following period from 02/04/2003 to 25/12/2006, as shown in Table ~\ref{TB:return-eigenvector-2}, four IT stocks and two electronics stocks are in the list of the largest ten components of $u^3$, and five machinery stocks and two petrochemicals stocks are in the list of $u^4$. More interestingly, stocks 600198, 600100, and 600770, which are among the largest ten components of $u^2$ in the first time period, appear in the largest ten components of $u^3$ in the following period. The starting dates of the moving windows in the first period are from 04/01/1999 to 30/07/2001, and from 30/07/2001 to 10/05/2005 for the second period. The evolutions of the IT and electronic industries recall the history of the Chinese stock market in the period of 1999-2001. During that period of time, the Chinese stock market was in a bull market, and high-tech stocks issued by companies deal in IT and electronics were leading the rise. After 2001, the Chinese stock market started to decline, thus the IT and electronics stocks are contained in $u^3$. Similar phenomenon is observed for the stocks in machinery and petrochemicals industries: stocks 600843, 600818, 600618, and 600841 among the largest ten components of $u^3$ in the first time period become the members of the largest ten components of $u^4$ in the following period.

The dynamic evolution behavior is also observed in real estate industry. In the period from 02/04/2003 to 25/12/2006, five stocks belonging to real estate industry appear in the largest ten components of $u^5$. The number of real estate stocks in the largest ten components of $u^4$ increases to seven in the period from 25/12/2006 to 13/01/2009. In the following period from 13/01/2009 to 11/05/2010, five (seven) real estate stocks are in the largest ten components of $u^3$ ($u^4$). After September 2008, the Chinese stock market tended to be affected by the global financial crisis, and the stocks belonging to real estate industry were leading the drop. Consequently, we observe that seven real estate stocks appear in the largest ten components of $u^2$ for the period from 11/05/2010 to 30/12/2011, in which the moving windows have starting dates from 16/09/2008 to 12/05/2010. In general, the real estate stocks contained in the largest five eigenvectors slowly move to be contained in the second largest eigenvector as the time approaches the global financial crisis. This conclusion is based upon the fact that many real estate stocks appear repeatedly in the largest ten components of the largest five eigenvectors in different periods. For instance, stock 600663 first appears in the largest ten components of $u^5$ in the period from 02/04/2003 to 25/12/2006, then it moves to be in those of $u^4$ in the following period from 25/12/2006 to 13/01/2009, and finally it becomes a member of those of $u^2$ in the latest period from 11/05/2010 to 30/12/2011.

\begin{table}[htp]
 \centering
 \caption{Largest ten components of $u^2$, $u^3$, $u^4$, and $u^5$ by the average ranks of the eigenvector components taken over the moving windows with ending dates from 06/09/2000 to 02/04/2003. The eigenvectors are obtained from the correlation matrices of the return series in these moving windows. The stock codes corresponding to the largest ten components, the industries they belonging to, and the industry codes are listed. } \label{TB:return-eigenvector-1}
\resizebox{8.8cm}{!}{ %
\begin{tabular}{lp{2.5cm}cllp{2.5cm}c}
  \hline
  \multicolumn{3}{c}{$u^2$} & & \multicolumn{3}{c}{$u^3$}\\  %
  \cline{1-3} \cline{5-7}
  Stock code & Industry & Industry code & & Stock code & Industry & Industry code \\
  \hline
    $600718$ & IT              & G  & & $600613$ & Pharmaceuticals            & C8 \\
    $600098$ & Utilities       & D  & & $600845$ & IT                         & G \\
    $600832$ & Comprehensive   & M  & & $600614$ & Real estate                & J \\
    $600198$ & IT              & G  & & $600822$ & Wholesale \& retail trade & H \\
    $600657$ & Real estate     & J  & & $600843$ & Machinery                  & C7 \\
    $600637$ & Electronics     & C5 & & $600619$ & Machinery                  & C7 \\
    $600100$ & IT              & G  & & $600818$ & Machinery                  & C7 \\
    $600776$ & IT              & G  & & $600618$ & Petrochemicals             & C4 \\
    $600138$ & Social services & K  & & $600841$ & Machinery                  & C7 \\
    $600770$ & IT              & G  & & $600688$ & Petrochemicals             & C4 \\
  \hline
  \multicolumn{3}{c}{$u^4$} & & \multicolumn{3}{c}{$u^5$}\\  %
  \cline{1-3} \cline{5-7}
  Stock code & Industry & Industry code & & Stock code & Industry & Industry code \\
  \hline
    $600623$ & Petrochemicals             & C4  & & $600773$ & Real estate                & J \\
    $600695$ & Food \& beverage           & C0  & & $600847$ & Machinery                  & C7 \\
    $600618$ & Petrochemicals             & C4  & & $600818$ & Machinery                  & C7 \\
    $600614$ & Real estate                & J   & & $600647$ & Real estate                & J \\
    $600613$ & Pharmaceuticals            & C8  & & $600696$ & Real estate                & J \\
    $600612$ & Other manufacturing        & C99 & & $600770$ & IT                         & G \\
    $600886$ & Utilities                  & D   & & $600058$ & Wholesale \& retail trade & H \\
    $600821$ & Wholesale \& retail trade  & H   & & $600608$ & Metals \& non-metals       & C6 \\
    $600079$ & Pharmaceuticals            & C8  & & $600055$ & Machinery                  & C7 \\
    $600841$ & Machinery                  & C7  & & $600792$ & Petrochemicals             & C4 \\
  \hline
\end{tabular} }%
\end{table}

\begin{table}[htp]
 \centering
 \caption{Largest ten components of $u^2$, $u^3$, $u^4$, and $u^5$ by the average ranks of the eigenvector components taken over the moving windows with ending dates from 02/04/2003 to 25/12/2006. The eigenvectors are obtained from the correlation matrices of the return series in these moving windows. The stock codes corresponding to the largest ten components, the industries they belonging to, and the industry codes are listed.} \label{TB:return-eigenvector-2}
\resizebox{8.8cm}{!}{ %
\begin{tabular}{lp{2.5cm}cllp{2.5cm}c}
  \hline
  \multicolumn{3}{c}{$u^2$} & & \multicolumn{3}{c}{$u^3$}\\  %
  \cline{1-3} \cline{5-7}
  Stock code & Industry & Industry code & & Stock code & Industry & Industry code \\
  \hline
    $600123$ & Mining          & B  & & $600171$ & Electronics                         & C5 \\
    $600009$ & Transportation  & F  & & $600602$ & Electronics                         & C5 \\
    $600848$ & Machinery       & C7 & & $600100$ & IT                                  & G \\
    $600098$ & Utilities       & D  & & $600088$ & Communication \& cultural industry  & L \\
    $600695$ & Food \& beverage& C0 & & $600832$ & Comprehensive                       & M \\
    $600740$ & Petrochemicals  & C4 & & $600775$ & IT                                  & G \\
    $600642$ & Utilities       & D  & & $600770$ & IT                                  & G \\
    $600795$ & Utilities       & D  & & $600624$ & Comprehensive                       & M \\
    $600096$ & Petrochemicals  & C4 & & $600198$ & IT                                  & G \\
    $600649$ & Comprehensive   & M  & & $600608$ & Metals \& non-metals                & C6 \\
  \hline
  \multicolumn{3}{c}{$u^4$} & & \multicolumn{3}{c}{$u^5$}\\  %
  \cline{1-3} \cline{5-7}
  Stock code & Industry & Industry code & & Stock code & Industry & Industry code \\
  \hline
    $600841$ & Machinery                  & C7  & & $600648$ & Wholesale \& retail trade  & H \\
    $600818$ & Machinery                  & C7  & & $600136$ & Real estate                & J \\
    $600612$ & Other manufacturing        & C99 & & $600823$ & Real estate                & J \\
    $600614$ & Real estate                & J   & & $600620$ & Real estate                & J \\
    $600623$ & Petrochemicals             & C4  & & $600781$ & Pharmaceuticals            & C8 \\
    $600822$ & Wholesale \& retail trade  & H   & & $600807$ & Real estate                & J \\
    $600843$ & Machinery                  & C7  & & $600663$ & Real estate                & J\\
    $600610$ & Machinery                  & C7  & & $600086$ & Other manufacturing        & C99 \\
    $600618$ & Petrochemicals             & C4  & & $600054$ & Social services            & K \\
    $600604$ & Machinery                  & C7  & & $600715$ & Machinery                  & C7 \\
  \hline
\end{tabular} }%
\end{table}

\begin{table}[htp]
 \centering
 \caption{Largest ten components of $u^2$, $u^3$, $u^4$, and $u^5$ by the average ranks of the eigenvector components taken over the moving windows with ending dates from 25/12/2006 to 13/01/2009. The eigenvectors are obtained from the correlation matrices of the return series in these moving windows. The stock codes corresponding to the largest ten components, the industries they belonging to, and the industry codes are listed.} \label{TB:return-eigenvector-3}
\resizebox{8.8cm}{!}{ %
\begin{tabular}{lp{2.5cm}cllp{2.5cm}c}
  \hline
  \multicolumn{3}{c}{$u^2$} & & \multicolumn{3}{c}{$u^3$}\\  %
  \cline{1-3} \cline{5-7}
  Stock code & Industry & Industry code & & Stock code & Industry & Industry code \\
  \hline
    $600660$ & Metals \& non-metals      & C6 & & $600610$ & Machinery             & C7 \\
    $600716$ & Real estate               & J  & & $600751$ & Transportation        & F \\
    $600773$ & Real estate               & J  & & $600711$ & Mining                & B \\
    $600809$ & Food \& beverage          & C0 & & $600733$ & Real estate           & J \\
    $600600$ & Food \& beverage          & C0 & & $600757$ & Textiles \& apparel   & C1 \\
    $600096$ & Petrochemicals            & C4 & & $600695$ & Food \& beverage      & C0 \\
    $600761$ & Machinery                 & C7 & & $600722$ & Petrochemicals        & C4 \\
    $600875$ & Machinery                 & C7 & & $600101$ & Utilities             & D \\
    $600887$ & Food \& beverage          & C0 & & $600664$ & Pharmaceuticals       & C8 \\
    $600694$ & Wholesale \& retail trade & H  & & $600608$ & Metals \& non-metals  & C6 \\
  \hline
  \multicolumn{3}{c}{$u^4$} & & \multicolumn{3}{c}{$u^5$}\\  %
  \cline{1-3} \cline{5-7}
  Stock code & Industry & Industry code & & Stock code & Industry & Industry code \\
  \hline
    $600663$ & Real estate                & J  & & $600691$ & Metals \& non-metals & C6 \\
    $600639$ & Real estate                & J  & & $600695$ & Food \& beverage     & C0 \\
    $600675$ & Real estate                & J  & & $600699$ & Petrochemicals       & C4 \\
    $600648$ & Wholesale \& retail trade  & H  & & $600133$ & Construction         & E \\
    $600638$ & Real estate                & J  & & $600724$ & Real estate          & J \\
    $600694$ & Wholesale \& retail trade  & H  & & $600634$ & Real estate          & J \\
    $600665$ & Real estate                & J  & & $600191$ & Food \& beverage     & C0\\
    $600622$ & Real estate                & J  & & $600884$ & Textiles \& apparel  & C1 \\
    $600732$ & Real estate                & J  & & $600757$ & Textiles \& apparel  & C1 \\
    $600858$ & Wholesale \& retail trade  & H  & & $600868$ & Utilities            & D \\
  \hline
\end{tabular} }%
\end{table}

\begin{table}[htp]
 \centering
 \caption{Largest ten components of $u^2$, $u^3$, $u^4$, and $u^5$ by the average ranks of the eigenvector components taken over the moving windows with ending dates from 13/01/2009 to 11/05/2010. The eigenvectors are obtained from the correlation matrices of the return series in these moving windows. The stock codes corresponding to the largest ten components, the industries they belonging to, and the industry codes are listed.} \label{TB:return-eigenvector-4}
\resizebox{8.8cm}{!}{ %
\begin{tabular}{lp{2.5cm}cllp{2.5cm}c}
  \hline
  \multicolumn{3}{c}{$u^2$} & & \multicolumn{3}{c}{$u^3$}\\  %
  \cline{1-3} \cline{5-7}
  Stock code & Industry & Industry code & & Stock code & Industry & Industry code \\
  \hline
    $600728$ & IT                         & G  & & $600648$ & Wholesale \& retail trade  & H \\
    $600890$ & Real estate                & J  & & $600732$ & Real estate                & J \\
    $600076$ & IT                         & G  & & $600620$ & Real estate                & J \\
    $600751$ & Transportation             & F  & & $600621$ & Electronics                & C5 \\
    $600891$ & Wholesale \& retail trade  & H  & & $600622$ & Real estate                & J \\
    $600773$ & Real estate                & J  & & $600062$ & Pharmaceuticals            & C8 \\
    $600892$ & Machinery                  & C7 & & $600694$ & Wholesale \& retail trade  & H \\
    $600757$ & Textiles \& apparel        & C1 & & $600716$ & Real estate                & J \\
    $600800$ & Comprehensive              & M  & & $600634$ & Real estate                & J \\
    $600714$ & Mining                     & B  & & $600750$ & Pharmaceuticals            & C8 \\
  \hline
  \multicolumn{3}{c}{$u^4$} & & \multicolumn{3}{c}{$u^5$}\\  %
  \cline{1-3} \cline{5-7}
  Stock code & Industry & Industry code & & Stock code & Industry & Industry code \\
  \hline
    $600067$ & Machinery             & C7 & & $600783$ & Metals \& non-metals        & C6\\
    $600675$ & Real estate           & J  & & $600836$ & Paper \& printing           & C3\\
    $600748$ & Real estate           & J  & & $600729$ & Wholesale \& retail trade   & H \\
    $600895$ & Real estate           & J  & & $600828$ & Wholesale \& retail trade   & H \\
    $600684$ & Real estate           & J  & & $600697$ & Wholesale \& retail trade   & H \\
    $600773$ & Real estate           & J  & & $600635$ & Utilities                   & D \\
    $600064$ & Real estate           & J  & & $600779$ & Food \& beverage            & C0\\
    $600109$ & Finance \& insurance  & I  & & $600887$ & Food \& beverage            & C0\\
    $600665$ & Real estate           & J  & & $600624$ & Comprehensive               & M \\
    $600084$ & Food \& beverage      & C0 & & $600106$ & Transportation              & F \\
  \hline
\end{tabular} }%
\end{table}

\begin{table}[htp]
 \centering
 \caption{Largest ten components of $u^2$, $u^3$, $u^4$, and $u^5$ by the average ranks of the eigenvector components taken over the moving windows with ending dates from 11/05/2010 to 30/12/2011. The eigenvectors are obtained from the correlation matrices of the return series in these moving windows. The stock codes corresponding to the largest ten components, the industries they belonging to, and the industry codes are listed.} \label{TB:return-eigenvector-5}
\resizebox{8.8cm}{!}{ %
\begin{tabular}{lp{2.5cm}cllp{2.5cm}c}
  \hline
  \multicolumn{3}{c}{$u^2$} & & \multicolumn{3}{c}{$u^3$}\\  %
  \cline{1-3} \cline{5-7}
  Stock code & Industry & Industry code & & Stock code & Industry & Industry code \\
  \hline
    $600748$ & Real estate                & J  & & $600058$ & Wholesale \& retail trade  & H \\
    $600067$ & Machinery                  & C7 & & $600757$ & Textiles \& apparel        & C1 \\
    $600823$ & IT                         & J  & & $600188$ & Mining                     & B \\
    $600675$ & Real estate                & J  & & $600838$ & Wholesale \& retail trade  & H \\
    $600657$ & Real estate                & J  & & $600117$ & Metals \& non-metals       & C6 \\
    $600664$ & Pharmaceuticals            & C8 & & $600692$ & Transportation             & F \\
    $600639$ & Real estate                & J  & & $600699$ & Petrochemicals             & C4 \\
    $600052$ & Real estate                & J  & & $600675$ & Real estate                & J \\
    $600663$ & Real estate                & J  & & $600606$ & Real estate                & J \\
    $600102$ & Metals \& non-metals       & C6 & & $600631$ & Wholesale \& retail trade  & H \\
  \hline
  \multicolumn{3}{c}{$u^4$} & & \multicolumn{3}{c}{$u^5$}\\  %
  \cline{1-3} \cline{5-7}
  Stock code & Industry & Industry code & & Stock code & Industry & Industry code \\
  \hline
    $600757$ & Textiles \& apparel        & C1 & & $600680$ & IT                         & G \\
    $600751$ & Transportation             & F  & & $600809$ & Food \& beverage           & C0 \\
    $600180$ & Agriculture                & A  & & $600651$ & Machinery                  & C7 \\
    $600102$ & Metals \& non-metals       & C6 & & $600607$ & Wholesale \& retail trade  & H \\
    $600891$ & Wholesale \&  retail trade & H  & & $600756$ & IT                         & G \\
    $600608$ & Metals \& non-metals       & C6 & & $600666$ & Pharmaceuticals            & C8 \\
    $600123$ & Mining                     & B  & & $600640$ & Wholesale \& retail trade  & H\\
    $600604$ & Machinery                  & C7 & & $600085$ & Pharmaceuticals            & C8 \\
    $600188$ & Mining                     & B  & & $600812$ & Pharmaceuticals            & C8 \\
    $600699$ & Petrochemicals             & C4 & & $600654$ & Electronics                & C5 \\
  \hline
\end{tabular} }%
\end{table}

\section{Conclusion}

In conclusion, we have conducted a thorough study of the evolution of the cross-correlations between the return series of 367 A-share stocks on Shanghai Stock Exchange from 1999 to 2011. We find that the stock returns behave more collectively in volatile periods, showing biased distribution of correlation coefficients centered around lager positive coefficients and larger values of mean correlation coefficient as the time approaches the two big crashes in 2001 and 2008. In the same volatile periods, we find that the largest eigenvalue shows larger values, while the number of eigenvalues that significantly deviate from those of the random correlation matrix is smaller. In addition, only a small number of eigenvalues can explain the major portion of the total system variance when the market is volatile, which indicates a high level of systemic risk.

We have further analyzed the components of the deviating eigenvectors and their contributions. By computing the contributions of the components grouped in conventional industries, we find that significant contributors, such as mining, electronics, IT, and real estate, for distinct eigenvectors over different $t$. We also analyze the projection of the stock returns on the largest eigenvector, and confirm the market-wide influence of the largest eigenvector and its stability in time. In the analysis of the component stocks which significantly contribute to each eigenvector, remarkable dynamic evolutions of conventional industries are observed, basically consistent with the results of industry contributions. The stocks in IT and electronics industries significantly contributing to the second largest eigenvector before the crash in 2001 become the significant contributors of the third largest eigenvector after the crash. Similarly, the stocks in real estate industry significantly contributing to other deviating eigenvectors before the crisis of 2008-2009 become the significant contributors of the second largest eigenvector during the crisis period.

\begin{acknowledgments}
 We thank M. Tumminello for helpful discussions. This work was partially supported by the National Natural Science Foundation (Nos. 10905023, 11075054 and 71131007), Humanities and Social Sciences Fund sponsored by Ministry of Education of the People's Republic of China (No. 09YJCZH042), the Shanghai (Follow-up) Rising Star Program Grant 11QH1400800, the Zhejiang Provincial Natural Science Foundation of China (Nos. Z6090130 and Y6110687), and the Fundamental Research Funds for the Central Universities.
\end{acknowledgments}

\bibliography{E:/Papers/Auxiliary/Bibliography}

\end{document}